\documentclass[journal=jacsat,manuscript=article,layout=twocolumn,hyperref]{achemso}
\usepackage{iftex,comment}
\usepackage{amsmath,amssymb,amsfonts} 
\usepackage{physics}
\usepackage[version=4]{mhchem}
\usepackage[super]{nth}
\usepackage{xfrac,calc,multirow}
\usepackage{nameref,doi,lineno,comment}
\usepackage{color,graphicx,titlesec}

\ifXeTeX%
\usepackage{mathspec}
\setallmainfonts[%
SizeFeatures={%
	{Size={-8.39},Font=* Caption},
	{Size={8.4-10.99},Font=* SmText},
	{Size={11-13.99},Font=* },
	{Size={14-21.49},Font=* Subhead},
	{Size={21.5-},Font=* Display}
},
BoldFont={* Bold},
BoldItalicFont={* Bold Italic},
]{Arno Pro}
\usepackage{polyglossia}
\setmainlanguage{english}
\else
\usepackage[english]{babel}
\usepackage{mathptmx}
\fi

\definecolor{JCTCGreen}{rgb}{0,0.48,0.25}
\definecolor{JACSBlue}{rgb}{0.15,0.2,0.5}
\definecolor{JACSYellow}{rgb}{1,0.8,0.2}
\definecolor{JPCBBlue}{rgb}{0.13,0.31,0.66}

\usepackage{doi,nameref,hyperref,zref-xr}
\zxrsetup{toltxlabel}
\zexternaldocument*[si:]{charged-si}
\usepackage[noabbrev,capitalize]{cleveref}
\crefname{equation}{eq~}{eqs~}     % abbreviation without a dot
\creflabelformat{equation}{#2#1#3} % no parenthesis
\newcommand{\crefx}[2]{\cref{#1}\,(#2)}

\setkeys{acs}{articletitle=true}
\setkeys{acs}{doi=true}

\newcommand{\figwidth}{\ifdim\linewidth>4in .6667\else 1\fi} % reasonable size in a single column mode
\graphicspath{{figs/}}

\usepackage[separate-uncertainty=false]{siunitx}
\DeclareSIUnit\molar{\mole\per\cubic\deci\metre}
\DeclareSIUnit\Molar{\textsc{m}}
\DeclareSIUnit\Boltzmann{\mathit{k_\mathrm{B}T}}
\DeclareSIUnit\percent{\text{\%}}

\newcommand{\refSI}{SI}
\newcommand{\abbr}[1]{#1}

\newcommand*{\epsion}{\epsilon_\text{ion}}
\newcommand*{\epscc}{\epsilon_\text{cation}}
\newcommand*{\epsca}{\epsilon_\text{anion}}
\newcommand*{\epsmer}{\epsilon_\mathrm{mer}}
\newcommand*{\Rg}{R_\mathrm{g}}
\newcommand*{\kB}{k_\mathrm{B}}
\newcommand*{\diff}{\mathop{\mathrm{d}\!}}
\newcommand{\iout}{_\mathrm{out}}
\newcommand{\iin}{_\mathrm{in}}
\newcommand{\imer}{_\mathrm{mer}}
\newcommand{\ipol}{_\mathrm{pol}}
\newcommand{\virial}[2]{B_{#1}^\mathrm{#2}}
\newcommand{\virialx}[2]{B_{#1}^{#2}}
\newcommand{\pdvat}[3]{\left.\pdv{#1}{#2}\right|_{#3}}

\titleformat{name=\section}%
{\fontseries{sb}\bf\sffamily\color{JACSBlue}\uppercase}%
{\rule{0.75em}{0.75em}}{0.75em}{}

\titleformat{name=\section,numberless}%
{\fontseries{sb}\bf\sffamily\color{JACSBlue}\uppercase}%
{\rule{0.75em}{0.75em}}{0.75em}{}

\titleformat{name=\subsection}%
{\fontseries{sb}\bf\sffamily\color{JACSBlue}}%
{}{0em}{}

\title{Tuning the collapse transition of weakly charged polymers by ion-specific screening and adsorption}
\ifXeTeX%
\author{Richard Chudoba}
\affiliation{Institut für Physik, Humboldt-Universität zu Berlin, Newtonstraße~15, D-12489 Berlin, Germany}
\alsoaffiliation{Physikalisches Institut, Albert-Ludwigs Universität Freiburg, Hermann-Herder-Straße~3, D-79104 Freiburg im Breisgau, Germany}
\alsoaffiliation{Research Group Simulations of Energy Materials, Helmholtz-Zentrum Berlin, Hahn-Meitner-Platz~1, D-14109 Berlin, Germany}
\email{richard.chudoba@helmholtz-berlin.de}
\author{Jan Heyda}
\affiliation{Department of Physical Chemistry, University of Chemistry and Technology, Prague, Technická~5, CZ-16628 Praha, Czechia}
\email{jan.heyda@vscht.cz}
\author{Joachim Dzubiella}
\affiliation{Physikalisches Institut, Albert-Ludwigs Universität Freiburg, Hermann-Herder-Straße~3, D-79104 Freiburg im Breisgau, Germany}
\alsoaffiliation{Research Group Simulations of Energy Materials, Helmholtz-Zentrum Berlin, Hahn-Meitner-Platz~1, D-14109 Berlin, Germany}
\email{joachim.dzubiella@physik.uni-freiburg.de}
\else%
\author{Richard Chudoba}
\affiliation{Institut f\"ur Physik, Humboldt-Universit\"at zu Berlin, Newtonstra{\ss}e~15, D-12489 Berlin, Germany}
\alsoaffiliation{Physikalisches Institut, Albert-Ludwigs Universit\"at Freiburg, Hermann-Herder-Stra{\ss}e~3, D-79104 Freiburg im Breisgau, Germany}
\alsoaffiliation{Research Group Simulations of Energy Materials, Helmholtz-Zentrum Berlin, Hahn-Meitner-Platz~1, D-14109 Berlin, Germany}
\email{richard.chudoba@helmholtz-berlin.de}
\author{Jan Heyda}
\affiliation{Department of Physical Chemistry, University of Chemistry and Technology, Prague, Technick\'a~5, CZ-16628 Praha, Czechia}
\email{jan.heyda@vscht.cz}
\author{Joachim Dzubiella}
\affiliation{Physikalisches Institut, Albert-Ludwigs Universit\"at Freiburg, Hermann-Herder-Stra{\ss}e~3, D-79104 Freiburg im Breisgau, Germany}
\alsoaffiliation{Research Group Simulations of Energy Materials, Helmholtz-Zentrum Berlin, Hahn-Meitner-Platz~1, D-14109 Berlin, Germany}
\email{joachim.dzubiella@physik.uni-freiburg.de}
\fi

\hypersetup{
  pdftitle={Tuning the collapse transition of weakly charged polymers by ion-specific screening and adsorption},
  pdfauthor={Richard Chudoba, Jan Heyda, Joachim Dzubiella},
}

\renewcommand{\tocentryname}{\fontseries{sb}\bf\sffamily Graphical TOC Entry}
\begin{tocentry}
  \centering \includegraphics[scale=.72]{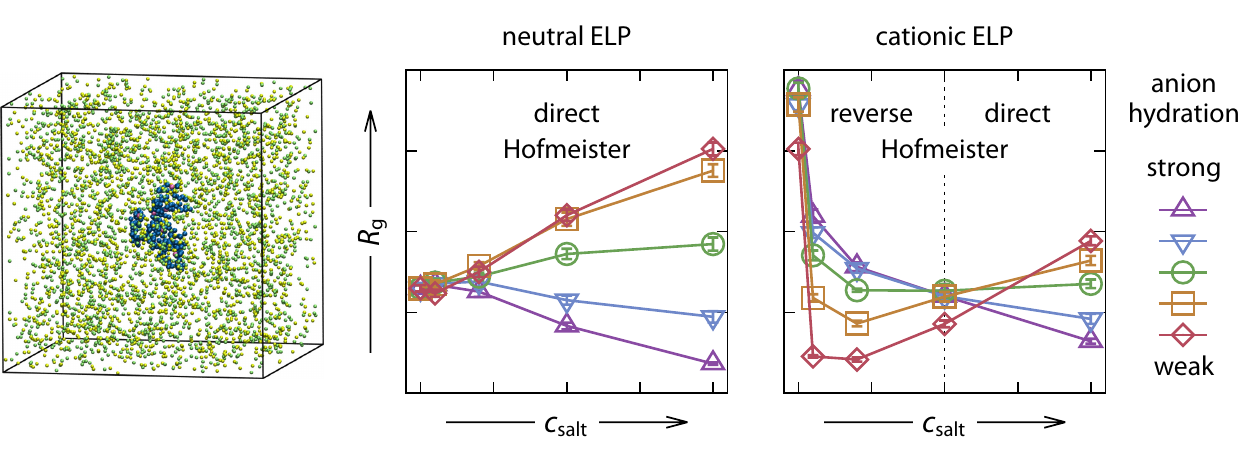}
\end{tocentry}

\begin{document}
	
  \renewcommand{\abstractname}{\uppercase{Abstract}}
  \begin{abstract}
    The experimentally observed swelling and collapse response of weakly charged polymers to the addition of specific salts displays quite convoluted behavior that is not easy to categorize. Here we use a minimalistic implicit solvent / explicit salt simulation model with a focus on ion-specific interactions between ions and a single weakly charged polyelectrolyte to qualitatively explain the observed effects.
In particular, we demonstrate ion-specific screening and bridging effects cause collapse at low salt concentrations whereas the same strong ion-specific direct interactions drive re-entrant swelling at high concentrations.
Consistently with experiments, a distinct salt concentration at which the salting-out power of anions inverts from the reverse to direct Hofmeister series is observed. At this, so called ‘isospheric point’, the ion-specific effects vanish.
Furthermore, with additional simplifying assumptions, an ion-specific mean-field model is developed for the collapse transition which quantitatively agrees with the simulations.
Our work demonstrates the sensitivity of the structural behavior of charged polymers to the addition of specific salt and shall be useful for further guidance of experiments.

  \end{abstract}
	
  \maketitle
	
  \section{Introduction}
\label{sec:introduction}

Biopolymers as well as functional synthetic polymers in solution are subject to a complex set of interactions that are decisive for their molecular structure and function, \abbr{e.g.}, as for folded versus unfolded states for proteins, or, more general, collapsed versus swollen states of a polymer.\cite{Dill2003} Typically those solvated polymers are weakly charged and constitute relatively heterogeneous, partially polar, partially hydrophobic, macromolecules. Hence their polymeric structure is greatly affected by the solvation environment. Apart from the solvent type (\abbr{e.g.}, water), the polymer conformation can be conveniently controlled by the concentration and type of added cosolutes. In particular, ion-specific, also called Hofmeister effects\cite{Hofmeister1888, Kunz2004} of salts on biopolymeric structure and phase behavior have been studied and highlighted in the recent years in an overwhelming amount of literature.\cite{Zhang2009, Schwierz2013, Jungwirth2014, Kou2015, Okur2017}

%Apart from the pure solvent type (\abbr{e.g.}, water), one of the most important experimentally controlled parameter is the type and concentration of added salt. In particular, ion-specific, also called Hofmeister effects\cite{Hofmeister1888, Kunz2004} on biopolymeric structure and phase behavior have been studied and highlighted in the recent years in an overwhelming amount of literature.\cite{Zhang2009, Schwierz2013, Jungwirth2014, Kou2015, Okur2017}

The Hofmeister effects on the polymeric structure and the phase behavior are very complex and polymer specific. In order to increase the accessibility to the problem, often model systems are employed. For instance, the poly-\textit{N}-isopropylacrylamide (PNIPAM) homopolymer or elastin-like polypeptides (ELPs), which both exhibit hydrophobic collapse transitions at their respective lower critical solution temperatures (LCST).\cite{Kherb2012, Suwa1998, Kujawa2001} These collapse transitions are analogous to cold-denaturation of proteins, in which ion-specific interactions (Hofmeister effects) play an prominent role.

PNIPAM is neutral polymer with LCST at \SI{32}{\degreeCelsius}. Charge groups can be introduced into the polymer structure by copolymerization, typically with weak or strong acids or bases.\cite{Kawasaki1997, Sasaki1999, Kawasaki2000} The electrostatic repulsion between charged monomers then leads to an increase of LCST in pure water and a strong response to added salt at millimolar salt concentration when charge screening sets in.

ELPs constitute a particularly nice model system as they can be expressed by bacteria in a strictly monodisperse fashion with different repeating sequences.\cite{Meyer2002, Meyer2004, Cho2008, McDaniel2013} It is possible, for instance, to synthesize pentapeptide sequences of the generic form \ce{(VPGXG)_n}, where X stays for any amino acid except proline.\cite{Urry1991, Urry1997} One can thus easily modulate the hydrophilic, hydrophobic, and charged character of the ELPs\cite{Urry1997} together with their LCST.\cite{McDaniel2013, Zhao2016} Salt-specific effects were studied not only for few neutral ELPs (\abbr{e.g.}, \ce{V}-120),\cite{Cho2008, Rembert2012, Heyda2017} but also for the weakly positively charged \ce{KV6}-120,\cite{Cho2009, Okur2017} and the weakly negatively charged \ce{DV2F}-64.\cite{Kherb2012}

%The cloud point temperature is often reported in the experiments as an approximate but direct measure of the LCST.\cite{Cho2009, Kherb2012, McDaniel}

\begin{figure}
	\centering
	\includegraphics[]{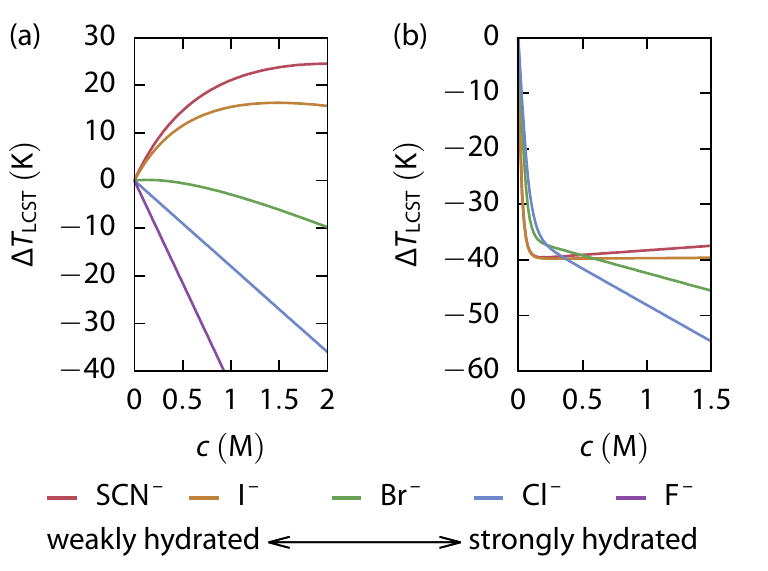}
	\caption{The change of the LCST of ELPs plotted versus the concentration of various sodium salts (with anions as given in the legend at the bottom) added to the polypeptide solution, as investigated experimentally.\cite{Cho2008, Cho2009} (a) Uncharged ELP \ce{V5A2G3-120}. (b) Weakly positively charged ELP \ce{KV6-112}.}
	\label{fig:experiment}
\end{figure}

For completely neutral polymers, \abbr{e.g.}, pure PNIPAM or neutral ELPs, only two regimes of salt action were reported in the literature,\cite{Cho2008, Zhang2005} see \crefx{fig:experiment}{a}. (I) Linear and rather steep decrease of the cloud point temperature (an approximate but direct measure of the LCST) with salt concentration was observed for most strongly hydrated salts, such as sulfates, acetates, fluorides, and chlorides. (II) Initial weak increase of the cloud point temperature at low salt concentrations, typically below \SI{1}{\Molar}, with subsequent decrease at higher concentrations. Only recently a third regime, which was predicted by computer simulations\cite{Heyda2013} and is characterized by strongly attractive bridging interaction, was found. Guanidinium thiocyanate, one of the most potent protein denaturants, steeply decreases the cloud point temperature below \SI{1.5}{\Molar}, and the temperature grows even more rapidly above \SI{1.5}{\Molar} salt.\cite{Heyda2017} The relation of this to the phenomenon of cononsolvency in solvent mixtures is under discussion.\cite{Budkov2018} Charged ELPs possess a more complex behavior due to the ion-specific screening at low salt concentrations,\cite{Cho2009, Okur2017, Kherb2012} cf.\ \crefx{fig:experiment}{b}, similar to the stability behavior of charged proteins.\cite{Zhang2009}

\enlargethispage{2pt}
It is now of fundamental importance to characterize the regimes in which specific salt-effects operate. This would allow to build a predictive model, which may be applied for determination of protein stability, but also for designing biomaterials of desired properties in their native environments. Computer simulations have already illuminated the action of varying solvents and cosolutes in coarse-grained simulations\cite{Winkler1998, Micka1999, Liu2002, Liu2003, Limbach2003, Jeon2007, Heyda2013, Lei2017} or all-atom simulations\cite{Algaer2011, Zhao2016, Backes2017, Mukherji2018}. Typically the results are interpreted by theories for chain structure and swelling,\cite{Dobrynin2005} \abbr{e.g.}, due to a counterion condensation at highly charged polymers\cite{Schiessel1998, Schiessel1999, Solis2000, Loh2008, Kundagrami2010} or specific steric interactions.\cite{Moncho-Jorda2013a, Moncho-Jorda2013b, Moncho-Jorda2014, Colla2014, Adroher-Benitez2015, Kosovan2015} Yet, a quantitative description of ion-specific counterion condensation relies on all-atom simulations.\cite{Heyda2012, Krishnamoorthy2018} Driven by increasing interest in cononsolvency and the action of mixed solutes on the polymer structure, a whole body of mean-field theories has been developed especially for the description of the coil-to-globule transition in mixed solvents.\cite{Heyda2013, Budkov2014, Budkov2014a, Budkov2014b, Schulz2015, Budkov2015, Budkov2016, Budkov2017, Budkov2018}

Recently, we have developed a thermodynamic approach for neutral polymers (such as PNIPAM, or neutral ELPs) within the framework of preferential binding,\cite{Heyda2014a, Okur2017} which helps interpreting the salt-specific thermodynamic fingerprints of cosolute-polymer interactions.\cite{Senske2016} A similar model was used and extended to include non-specific screening effects for weakly charged polymers.\cite{Heyda2014} In this work we aim to fully understand the complex data for charged elastin\cite{Cho2009, Okur2017, Kherb2012} (selected data shown in \crefx{fig:experiment}{b}). For this purpose we devise a generic simulation model of a charged polymer with different charge fractions in uni-univalent salt solutions. The ion-specificity of salts steps in by systematically varying the polymer-ion interactions, \abbr{i.e.}, from more repulsive, mimicking strongly hydrated ions (extreme `kosmotropes'), to very attractive, mimicking weakly hydrated ions (extreme `chaotropes').\cite{Kunz2004} Our previous simulations of neutral polymers exhibiting simple upper CST (UCST) have already demonstrated a non-trivial response of the polymer, where collapse by depletion, swelling through weak attraction, and re-entrant collapse by strong attractions have been observed.\cite{Heyda2013} As indicated by experiments, the situation becomes more complex in charged systems, where screening and synergistic cation and anion interactions with the polymer are expected. In this work, we therefore complement the simulations with a new theory that extends our Flory-type mean-field model for neutral polymers with approaches for ion-specific screening.\cite{Heyda2014, Moncho-Jorda2016}

  \section{Model and methods}
\label{sec:model}
In our model, a single polymer chain containing $N\imer=200$ coarse-grained monomer units is used. Monomers bearing a charge $+e$ or $-e$ are evenly (equidistantly) distributed along the chain to obtain the desired charge fractions $\xi$ of \SIlist[list-units=single]{0;5;10}{\percent}, for cationic or anionic chains, respectively.  Furthermore, coarse-grained ions (counter-ions of the polyelectrolyte and uni-univalent electrolyte) are explicitly resolved in the implicit solvent environment. All beads, \abbr{i.e.}, monomers and ions, interact via the Lennard-Jones (LJ) pair potential, that is
\begin{align}
\label{eq:lennard-jones}
U_{\mathrm{LJ}}^{\,ij} = 4\epsilon_{ij}\left[
\left(\frac{\sigma_{ij}}{r}\right)^{12}
- \left(\frac{\sigma_{ij}}{r}\right)^{6}
\right]\text.
\end{align}
The charged particles furthermore interact through the standard Coulomb potential. The relative permittivity of the background corresponds to water at room temperature with a dielectric constant $\epsilon_\mathrm{r} = 78$. This results in the Bjerrum length $l_\mathrm{B}= e^2 / \epsilon_\mathrm{r} k_\mathrm{B}T = \SI{0.7}{\nano\meter}$ at $T = \SI{298}{\kelvin}$, where $e$ is the elementary charge and $\kB$ is the Boltzmann constant. 

The bonds between adjacent monomers in the polymer chain are modeled with a harmonic potential with a spring constant $k=\SI{100}{\Boltzmann\per\nano\meter\squared}$ and a bond length $b=\SI{0.38}{\nano\meter}$.\cite{Heyda2013} %\cite{Toan2008}
Neither angular, nor torsional potential is applied. Non-bonded interactions between adjacent monomers are excluded.

The non-interacting polymer chain having $N\imer=200$ segments each $b=\SI{0.38}{\nano\meter}$ long is considered as an ideal reference of our simulation setup. The ideal radius of gyration equals to $\Rg^\mathrm{id} =  b \sqrt{N\imer/6} = \SI{2.19}{\nano\meter}$.

%\begin{align}
%\label{eq:coulomb}
%U_{ij}^\mathrm{\,C} = \frac{1}{4\pi\epsilon_0\epsilon_\mathrm{r}}\frac{q_iq_j}{r}
%\end{align}

The solvation characteristics of ions cannot be captured directly in the implicit solvent model. Instead the value of the second virial coefficient $B_2$ for the monomer--ion interaction (without electrostatics) can be used as a global measure of the hydrated character and ion-specific affinity to the polymer. Strongly hydrated ions prefer to stay in the bulk solution thus there is a net repulsion between them and the uncharged polymer. That corresponds to a positive $B_2$ value of the monomer-ion interaction. On the other hand, weakly hydrated ions tend to preferentially adsorb to the polymer surface, which gives a net negative $B_2$ value of the monomer-ion interaction. The value of $B_2$ can be computed from the pairwise potential (\cref{eq:b2-potential}). When the LJ potential is employed, $\epsilon \lesssim \SI{0.3}{\Boltzmann}$ gives a positive $B_2$ value (a net repulsion), while $\epsilon \gtrsim \SI{0.3}{\Boltzmann}$ yields to a negative $B_2$ value (a net attraction).

The implicit-solvent simulations are symmetric regarding the positive and negative charge. However, as known from experiments, anions exhibit greater variety of interactions with ELPs than common cations.\cite{Okur2017} To investigate a broader range of interaction types and yet to stay consistent with the experiments, we varied the anion--polymer interaction strength while keeping the strength of cation--polymer interaction unchanged. The affinity of the ions to the polymer is controlled by $\epscc$ and $\epsca$ in the LJ potential acting between a monomer (regardless of its charge) and a cation, or an anion, respectively. The value of $\epscc = \SI{0.1}{\Boltzmann}$ is kept constant, corresponding to strongly hydrated cations. The value of $\epsca$ varies between \SIrange[range-phrase={ and }]{0.1}{0.9}{\Boltzmann}, altering the value of the second virial coefficient of the interaction between anions and polymer. The actual $B_2$ values of the interactions employed in the simulations are summarized in \cref{si:tab:virial}.

%The importance of the $B_2$ value for modeling various ions in the Hofmeister series is discussed in detail later.

The non-electrostatic interaction between two ions (either anions, cations or cation-anion pairs) is determined by $\epsilon_\mathrm{ij} = \SI{0.3}{\Boltzmann}$ in all simulations, corresponding to the zero value of the second virial coefficient $B_2$ for a purely non-electrostatic interaction.

The non-bonded pair interaction between monomers reflects the quality of the implicit solvent towards the polymer, \abbr{i.e.}, varying solvent and polymer temperature. Our simple polymer model without solvent and explicit $T$-dependent interactions exhibits a UCST, while all qualitative trends including LCST are reflected, however, in inverse manner. LJ potentials with $\epsmer$ between \SIrange[range-phrase={ and }]{0.1}{0.9}{\Boltzmann} are used when scanning for the near-critical (collapse transition) solvent condition.

The value of $\sigma = \SI{0.3385}{\nano\meter}$ in the LJ potential is kept constant for all interactions,\cite{Heyda2013} except for the interaction between cation and monomer, where $\sigma = \SI{0.4787}{\nano\meter}$ corresponds to even more strongly hydrated cations interacting with rather apolar monomers. 

Langevin dynamics (LD) simulations in Gromacs 5.1\cite{Abraham2015} in $NVT$ ensemble are performed. The friction constant $\gamma$ is set to \SI{1.0}{\per\pico\second}. The stochastic 
processes correspond to the temperature $T = \SI{298}{\kelvin}$. The time step of the velocity Verlet integrator is set to \SI{7.5}{\femto\second}. Together with the reduced molar mass 
\SI{8.635}{\gram\per\mole} of all coarse-grained particles, the polymer conformations are sampled effectively. The electrostatic potential is computed using particle mesh Ewald method.\cite{Darden1993} To efficiently employ this method in a system with a low particle density, the value of a cut-off distance in the real space and the number of grid points in the Fourier reciprocal space are optimized, as described in \cref{si:tab:pme}.

A single coarse-grained polymer chain and salt ions are placed into a cubic simulation box to prepare \SI{0.0}{\Molar} (counterions only), \SIlist{0.1;0.4;1.0;2.0}{\Molar} salt solvation environment. Periodic boundary conditions are applied. The default volume of the simulation box \SI{8304}{\cubic\nano\meter} is reduced to one fourth when the polymer holds a compact globular conformation and the salt concentration is $\geq\SI{1}{\Molar}$. The compositions of the systems are described in detail in \cref{si:tab:pme}. A trajectory of \SI{525}{\nano\second} is obtained in every simulation run. The initial \SI{25}{\nano\second} is considered as an equilibration phase and thus excluded from the analysis. Standard Gromacs tools, namely gmx polystat and gmx rdf, are used for the data analysis. Further statistical evaluations are performed in R.\cite{RCoreTeam2015} The uncertainties of the quantities are determined as a standard error obtained from ten independent parts of the simulation trajectory.

\begin{figure}
	\centering
	\includegraphics[width=12em]{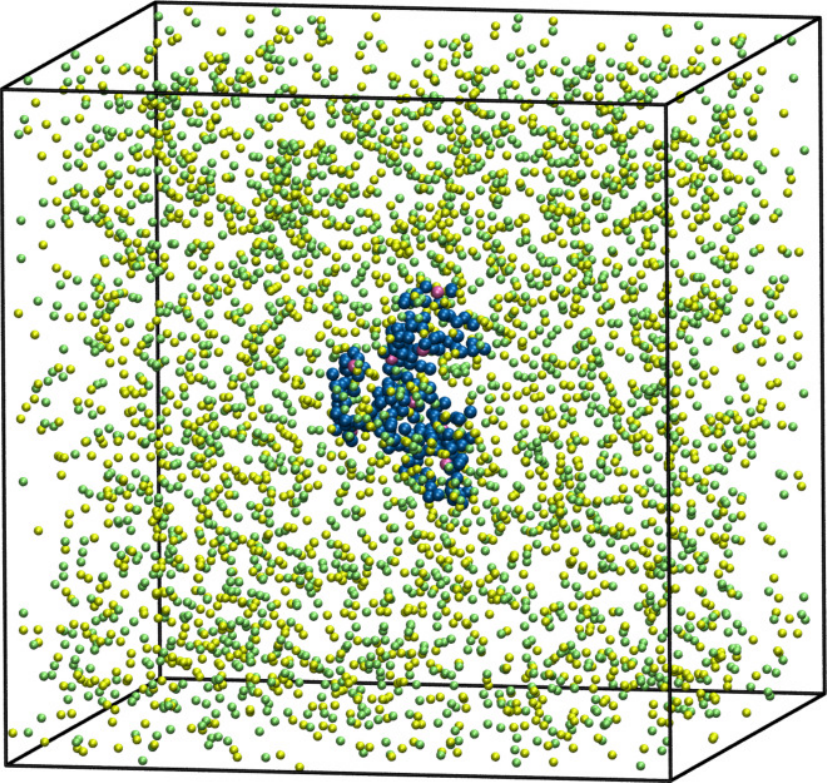}
	\label{fig:snapshot}
	\caption{A snapshot of the simulation box containing a 200-mer chain in \SI{0.4}{\Molar} salt solution within an implicit solvent. The neutral polymer beads are depicted in blue, while the charged ones in purple. Univalent cations and anions of the salt are presented as green and yellow beads, respectively.}
\end{figure}

  \section{Mean-field theory}
\label{sec:theory}

The change of the critical conditions of a polymer in a presence of cosolute can be described by a mean-field theory as follows. All interactions within the theory are described by a virial expansion and the electroneutrality condition is postulated. We choose an $NpT$ ensemble for the polymer which is in contact with a reservoir of cosolutes at a constant chemical potential. Cosolute molecules are allowed to exchange between the ensemble and the reservoir. At equilibrium the Gibbs free energy is minimized in such system. The polymer is simply modeled as an elastic, penetrable sphere that can change its radius upon the sorption of cosolutes. (Polymer conformational degrees of freedom are thus averaged out into an effective blob, like in the classical Flory approaches where only the polymer size is a variable.) Hence, the total Gibbs free energy of the polymer phase which is in a contact with a salt reservoir containing $i$ cosolute constituents, \abbr{i.e.}, anions and cations in our system, can be defined as
\begin{align}
G\ipol &= F\ipol + pV\ipol - \sum_i \bar\mu_i N_i\text{,}
\label{eq:free}
\end{align}
where $N_i$ is the number of particles of the respective cosolute constituent (in this work, only ions) contained within the polymer sphere of the effective volume $V\ipol$, $\bar\mu_i$ is the electrochemical potential of the respective constituent in the reservoir, $p$ is the osmotic pressure, and $F\ipol$ is the total Helmholtz free energy of the polymer including the absorbed cosolute.

The electrochemical potential $\bar\mu_i$ can be expressed with the explicit electrostatic term as 
\begin{align}
\bar\mu_i = \mu_i + z_i e \phi\text{,}
\label{eq:qchem}
\end{align}
where $\mu_i$ is a chemical potential of the $i$-th constituent, having a charge $z_i e$, which is located in an electrostatic field of a potential $\phi$. If the electroneutrality condition inside the effective volume of the polymer is assumed, the electrostatic part of the potential does not have to be considered, as no net charge is exchanged between the polymer spherical region and the reservoir. Thus we will use only the chemical potentials $\mu_i$ of the constituents. The electroneutrality is described with the condition
\begin{align}
\xi N\imer z\imer + \sum_i N_i z_i = 0\text{,}
\label{eq:qneutrality}
\end{align}
where $N\imer$ is the total number of monomers in a polymer chain, and $\xi$ is the fraction of charged segments, each bearing the charge $z\imer e$.

The interactions between particles are described by a virial expansion up to the fourth order in our model. Note, the expansion at least up to the third order is required to guarantee a physically relevant solution for the negative second virial coefficient. The second viral coefficient $\virialx2{ij}$ between the particles $i$ and $j$ can be easily evaluated from the Lennard-Jones pair potential $U^{ij}_\mathrm{LJ}$,
\begin{align}
\label{eq:b2-potential}
\virialx2{ij} = -2 \pi \int_0^\infty r^2 \left[\exp(-\beta U^{ij}_\mathrm{LJ})-1\right] \dd{r}\text{.}
\end{align}

We approximate all three body interactions using the third virial coefficient $\virialx3{ijk} = 2.00\,\sigma^6$, consistently with our previous work.\cite{Heyda2013} The four body interactions are modeled analogously with the fourth virial coefficient of hard spheres with a diameter $\sigma$, $\virialx4{ijkl} = 2.64\,\sigma^9$.

The polymer chain itself is modeled as an elastic (entropic)
spring\cite{Fixman1962, Budkov2014a}
\begin{align}
\beta F_\text{chain} = \frac49 (\alpha^2 + \alpha^{-2})\text{,}\quad
\alpha = R\ipol / (b \sqrt{\tfrac N6})\text{,}
\end{align}
where $\alpha$ is an elongation of the chain relative to the ideal chain having $N$ segments of a length $b$, and $\beta = (\kB T)^{-1}$ with the Boltzmann constant $\kB$ and temperature $T$. To relate the polymer radius $R\ipol$ and volume $V\ipol$, a spherical shape of the polymer is assumed. The total Helmholtz free energy of the polymer is then given as
\begin{align}
\begin{split}
\beta F\ipol = \beta F_\text{chain}
+ V\ipol\left[
\sum_i \rho_i \ln \rho_i +
\sum_{j,k} \virialx2{jk} \rho_j\rho_k + {}
\right.\\\left.
\frac12 \sum_{j,k,l} \virialx3{jkl} \rho_j\rho_k\rho_l +
\frac13 \sum_{j,k,l,m} \virialx4{jklm} \rho_j\rho_k\rho_l\rho_m
\right]\text{,}
\end{split}
\label{eq:fglob}
\end{align}
where $\rho$ is the particle density inside the effective polymer volume and the index $i$ stands for the cosolute constituents and the indices $j,k,l,m$ stand for the chain segments and/or the cosolute constituents.

The reservoir is modeled as a real gas which follows the fourth order virial expansion
%\begin{small}
\begin{align}
\begin{split}
\beta F\iout = V\iout\left[
2 \rho\iout \ln \rho\iout +
4\virial2{} \rho\iout^2 +
{}\vphantom{\frac{8\virial3{}}2}\right.\\
\left.
\frac{8\virial3{}}2 \rho\iout^3 +
\frac{16\virial4{}}3 \rho\iout^4
\right]\text{,}
\end{split}
\end{align}
%\end{small}
where $\rho\iout$ is the density of (uni-univalent) salt in the reservoir. The osmotic pressure $p$ and the chemical potentials $\mu_i$ in the reservoir can consequently be expressed as
\begin{align}
p &= \pdvat{F\iout}{V\iout}{N_j,T}\text{,}\\
\mu_i &= \pdvat{F\iout}{{N_i}}{N_{j \neq i},V,T}\text{,}
\end{align}
respectively, for any given cosolute concentration $\rho\iout$.

In order to find the equilibrium we numerically minimize the Gibbs free energy of the polymer $G\ipol$ (\cref{eq:free}) with respect to its effective volume $V\ipol$ and the number of cosolute particles $N_i$. Note that the theory can model both UCST and LCST polymers, depending on the input, in particular, $T$-dependence of the viral coefficients. 

Knowing the equilibrium concentration of the ions inside the polymer phase, the Donnan potential can be evaluated using \cref{eq:qchem} directly, for either cation, or anion. The chemical part $\mu_i$ of the electrochemical potential $\bar\mu_i$ of the respective cosolute $i$ inside the polymer phase can be expressed from \cref{eq:fglob}, computing the derivative
\begin{align}
\mu_{i,\mathrm{in}} &= \pdvat{F\ipol}{N_i}{N_{j \neq i},V\ipol,T}\text{.}
\label{eq:chemin}
\end{align}
Detailed evaluation of the Donnan potential is provided in the \refSI.

The mean effective volume of the polymer can be obtained as a Boltzmann weight average.
The fluctuations in the radius of the polymer $(\langle R\ipol^4\rangle / \langle R\ipol^2\rangle^2 -1)$ are computed as a function of the solvent quality ($\epsmer$). At the critical solution temperature (CST) the fluctuations maximize.\cite{Ivanov1998} The $n$-th moment of $R\ipol$ was computed using the Boltzmann distribution
\begin{align}
\label{eq:rg-boltzmann}
\langle R\ipol^n\rangle = \frac
{\iint_{R,\rho} R\ipol^n \exp(-\beta G\ipol) \diff \rho\iin \diff R\ipol}
{\iint_{R,\rho} \exp(-\beta G\ipol) \diff \rho\iin \diff R\ipol}\text{,}
\end{align}
where $\rho\iin$ is the salt concentration inside the polymer phase, maintaining the electroneutrality condition, and $R\ipol$ is the effective polymer radius ($V\ipol = 4/3 \pi R\ipol^3$).

Note that our mean-field theory can be considered valid only close to the coil-to-globule equilibrium when the polymer chain forms a rather compact conformation. The further assumptions are that the fraction of charged monomers has to be low and the salt concentration relatively high, \abbr{i.e.}, the neutral reference state is only slightly perturbed. 

A similar thermodynamic description has been recently developed for microgels, albeit a different procedure to find the thermodynamic equilibrium was employed.\cite{Moncho-Jorda2016} Salt-specific effects on weakly charged copolymers can be captured qualitatively by a simplified theory which treats the ion-specific and non-specific electrostatic contributions on a simple additive level, as demonstrated on PNIPAM copolymers.\cite{Heyda2014} In contrast, the theory developed in this work includes the ion-specific and non-specific interactions in a self-consistent way.

The equations are solved using numeric solvers of Wolfram Mathematica 11.2 software package.\cite{Mathematica2017}

  \section{Results and discussion}
\label{sec:results}

\subsection{Solvent quality near critical conditions}
\label{sec:results-solvent}

\begin{figure}[t]
	\centering
	\includegraphics[]{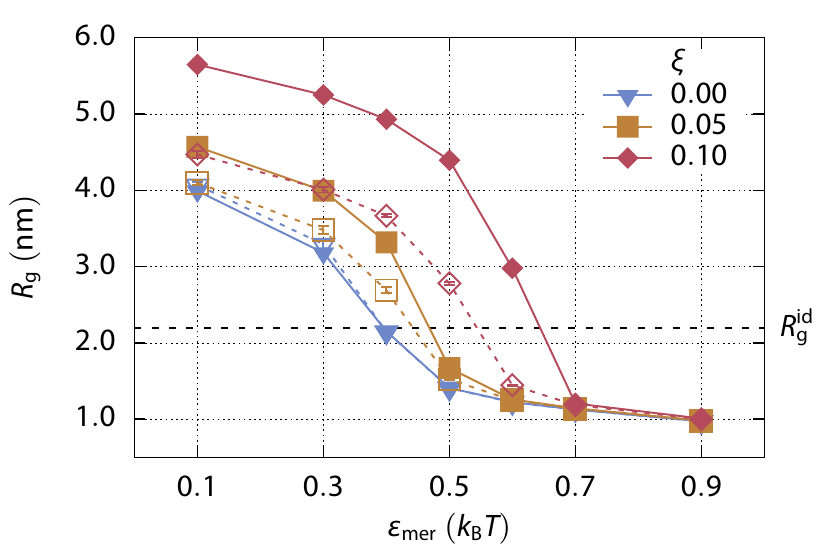}
	\caption{Mean radius of gyration $\Rg$ of a 200-mer as a function of the solvent quality, expressed by the monomer interaction strength $\epsmer$. Various charged fractions $\xi$ are distinguished by the symbol shape and color. Values for the chains in a pure solvent are shown with filled symbols while empty symbols are used for chains in $\SI{0.1}{\Molar}$ solution of a repulsive salt. Lines serve as a guide to the eye. The radius of gyration of the ideal chain $\Rg^\mathrm{id}$ is shown for comparison.}
	\label{fig:nosalt}
\end{figure}

In order to find conditions close to the swollen-to-collapse transition, a solvent quality parameter, \abbr{i.e.}, an effective interaction between monomers $\epsmer$, is varied from \SIrange{0.1}{0.9}{\Boltzmann}. Polymer chain conformation changes from an extended coil to a collapsed globule in this solvent quality range, as shown in \cref{fig:nosalt}.

%As a start, we scan over the solvent quality parameter $\epsmer$ to find the conditions close to the swollen-to-collapse transition.
%When the ELP is heated above the LCST, a phase separation occurs and a polymer-rich phase arises. Such increase in the polymer concentration corresponds to a swollen-to-collapse transition for a single chain, indicated by the sudden decrease in the radius of gyration. Thus the mean value of the radius of gyration is used as an indicator of the critical collapse condition, \abbr{i.e.}, as a simple measure of the CST or cloud point. The parameter $\epsmer$, describing the interaction between monomers, is gradually increased from \SIrange{0.1}{0.9}{\Boltzmann}. The polymer chain conformation changes from an extended coil to a collapsed globule in this solvent quality range, as shown in \cref{fig:nosalt}.

The radius of gyration of an uncharged polymer chain ($\xi=0.00$) at $\epsmer \simeq \SI{0.4}{\Boltzmann}$ almost exactly corresponds to the radius of gyration of the ideal, non-interacting polymer chain $\Rg^\mathrm{id}$. The radius of gyration increases with the charge fraction of the polymer and the transition is shifted to higher $\epsmer$ values. However, a small concentration of salt (\SI{0.1}{\Molar}) already suppresses this trend due to the charge screening, as shown in \cref{fig:nosalt}.
%Thus we choose $\epsmer = \SI{0.4}{\Boltzmann}$ to describe the solvent quality for uncharged and cationic polymer close to the critical conditions. For the strongly charged ($\xi=0.10$) anionic polymer, we use $\epsmer = \SI{0.5}{\Boltzmann}$ instead.
Thus we choose $\epsmer = \SI{0.4}{\Boltzmann}$ to describe the solvent quality for uncharged and charged polymers close to the critical conditions, with the exception of strongly charged ($\xi=0.10$) anionic polymer where the value $\epsmer = \SI{0.5}{\Boltzmann}$ is used instead.
Note that the critical value of $\epsmer$ depends on the actual chain size and the charge fraction. The value approaches \SI{0.3}{\Boltzmann} for an infinitely long uncharged chain.

%We perform implicit solvent simulations to investigate the effect of various salt types on both uncharged and charged polymer chain. As we focus on the CST transition, after the initial scan (SI) over the solvent quality parameter $\epsmer$, we choose $\epsmer = \SI{0.4}{\Boltzmann}$ to be close to the coil-globule equilibrium. The radius of gyration of a bare uncharged polymer chain under these conditions almost exactly corresponds to the radius of gyration of ideal, non-interacting polymer chain.

\subsection{Salt interaction with an uncharged ELP}
\label{sec:results-uncharged}

\begin{figure*}[t!]
	\centering
	\includegraphics[]{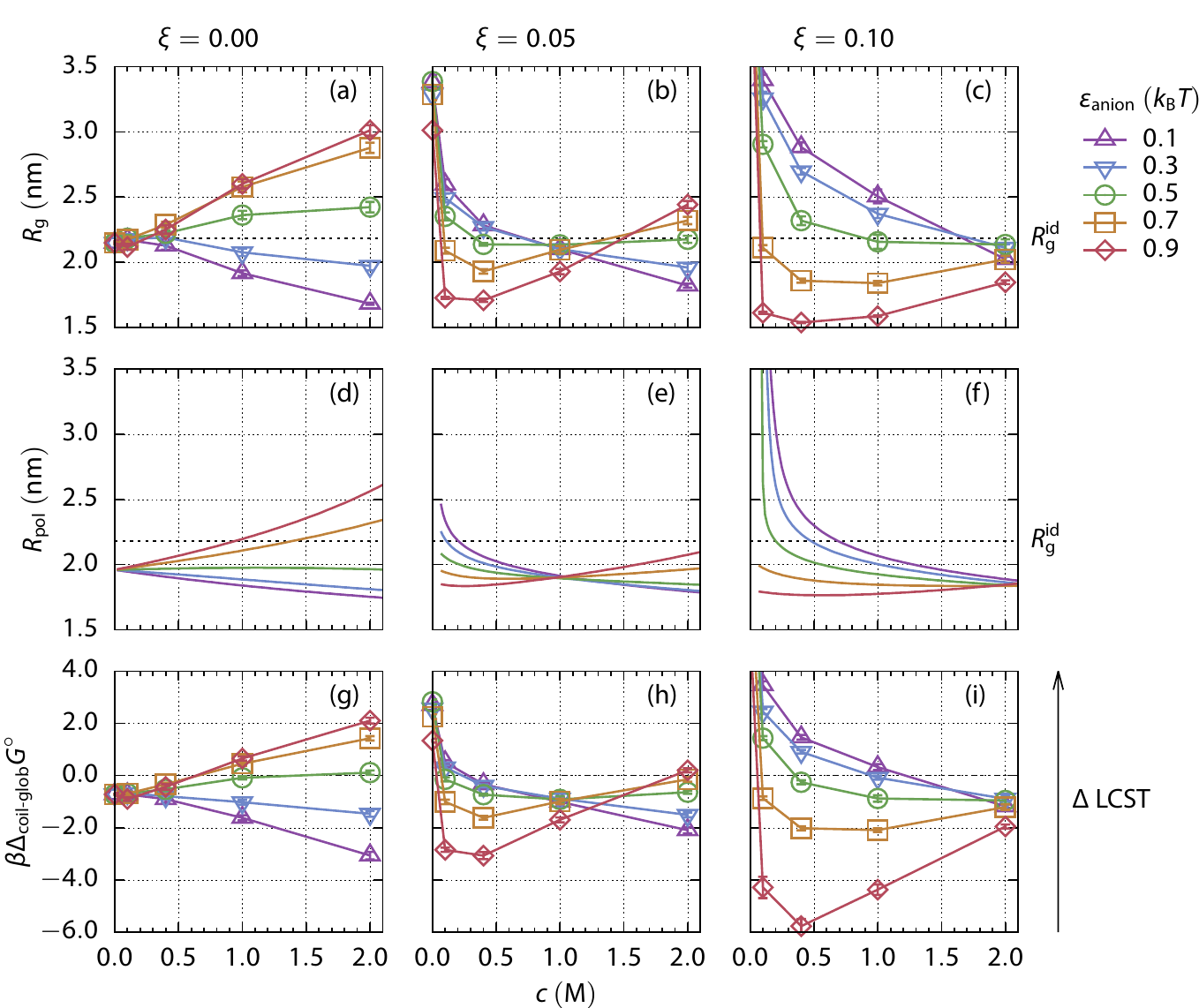}
	\caption{(a--c) Mean radius of gyration $R_\mathrm{g}$ obtained from simulations of 200-mer as a function of salt concentration. The polymer bears various charge fractions $\xi$ of positively charged monomers: (a) $0.00$, (b) $0.05$, and (c) $0.10$, respectively. Symbols and colors in the plots indicate the various strength of the interaction between the anions and the monomers, $\epsca$. Lines serve as a guide for the eye.
		%Marks on the x-axes represent the simulated concentrations.
		The mean radius of gyration of an ideal chain $\Rg^\text{id}$ is indicated. (d--f) An effective radius of the polymer $R\ipol$ computed using the mean-field theory for the very same systems (column-wise). The same color code is used for the strength of the anion-polymer interaction. (g--i) The standard free energy of the coil-to-globule transition, $\Delta_\text{coil-glob} G^\circ$, computed from the simulations. An increase in the energy of the transition corresponds to the elevation of LCST, proportional to $\Delta T_\mathrm{LCST}$ in \cref{fig:experiment}.}
	\label{fig:sim-thermodynamics}
	\label{fig:theory-thermodynamics}
\end{figure*}

\begin{figure}
	\centering
	\includegraphics[]{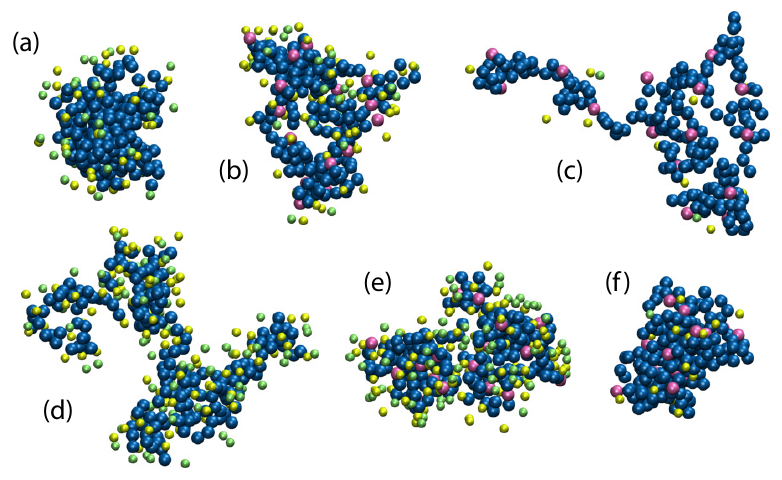}
	\caption{
		Simulation snapshots of typical polymer chain conformations under various conditions: uncharged 200-mer in \SI{2}{\Molar} salt solution (a, d), highly charged 200-mer in \SI{2}{\Molar} (b, e) and \SI{0.1}{\Molar} (c, f) salt solution. The salt consists of a very repulsive cation and a repulsive (a--c), or attractive (d--f) anion, respectively. Neutral and positively charged monomers are depicted in blue and purple, respectively. Green cations and yellow anions are are shown within a distance of \SI{0.7}{\nano\meter} from the polymer chain.
	}
	\label{fig:snapshots}	
\end{figure}

First, we investigate the uncharged ELP and relate qualitatively the simulation results to the known experimental data. Two different regimes, determined by the salt type, can be observed when the mean radius of gyration of the uncharged 200-mer is plotted as a function of the salt concentration, \crefx{fig:sim-thermodynamics}{a}. The repulsive salts ($\epsca = \text{\SIlist{0.1;0.3}{\Boltzmann}}$) induce a collapse through repulsion as the salt concentration increases. The salts with an attractive anion ($\epsca = \text{\SIlist{0.7;0.9}{\Boltzmann}}$) cause swelling of the polymer coil by attraction. Both regimes are described in detail for  non-ionic cosolutes in our previous work,\cite{Heyda2013} with the typical conformations of the polymer chain illustrated in \crefx{fig:snapshots}{a,\,d}, respectively.

The change in the radius of gyration can be also interpreted as a change of the standard free energy of the coil-to-globule transition
\begin{align}
\Delta_\text{coil-glob}G^\circ = - \kB T \ln \frac{p_\text{glob}}{p_\text{coil}}\text{,}
\end{align}
where $p$ is the probability of the respective state. The probabilities can be determined from the distribution of the radius of gyration. The value of the ideal radius of gyration can be used as a reasonable dividing boundary between the collapse and swollen states.  As we see, the standard free energy of the coil-to-globule transition changes when the salt is added to the polymer, \crefx{fig:sim-thermodynamics}{g}. The shift in the free energy has to be compensated by the change in the quality of the solvent to maintain the critical conditions. The quality of the solvent has to improve as the free energy decreases to reestablish the population of extended states, and vice versa. For polymers exhibiting a LCST, like ELPs, a decrease in the free energy of the coil-to-globule transition manifests as an LCST decrease. Polymers exhibiting an UCST, as our simple polymer model, show the opposite trend at the critical temperature.

The standard free energy of the coil-to-globule transition of the uncharged polymer obtained from simulation \crefx{fig:sim-thermodynamics}{d} can be qualitatively compared with the experimental LCST data of uncharged ELP, as shown in \crefx{fig:experiment}{a}. The anion transients from a strongly hydrated (little polymer affinity) to a weakly hydrated (large polymer affinity) as the value of $\epsca$ increases. Comparing to the charged polymer discussed later, only ion-specific effects can be observed both in the simulation and in the experiment.

Thiocyanate \ce{SCN-} is a weakly hydrated anion thus relatively high $\epsca \sim \SI{0.9}{\Boltzmann}$ is its appropriate parametrization in our simple model. The simulations predict a shift towards the swollen state of the uncharged polymer as the concentration of the salt increases, which agrees with the LCST elevation in the experimental results. Iodide \ce{I-} is another weakly hydrated anion (from the observed behavior $\epsca \sim \SIrange{0.5}{0.7}{\Boltzmann}$) having the same but weaker effect on uncharged ELP both in simulations and experiments. Bromide has the slightest effect of the salts experimentally studied and the anion with $\epsca \sim \SI{0.3}{\Boltzmann}$ exhibits the similar effects in the simulation. Moving towards the strongly hydrated anions in the Hofmeister series, chloride \ce{Cl-} and fluoride \ce{F-}, the LCST drops significantly in the experiments. The same transition is observed in the simulations containing anions with $\epsca \sim \text{\SIrange{0.1}{0.3}{\Boltzmann}}$. It can thus be concluded from the simulations and the experimental data that the salting-out (globule stabilization) power of anions follows the direct Hofmeister series for the uncharged ELP.

\subsection{Salt interaction with a positively charged ELP}
\label{sec:results-cationic}

Now we turn to a cationic ELP. For this, a polymer chain containing 200 monomers of which 20 are positively charged (a charge fraction $\xi=0.10$) is simulated. Its radius of gyration $\Rg$ and the standard free energy of the coil-to-globule transition as a function of salt concentration are shown in \crefx{fig:sim-thermodynamics}{c,\,i}. Two distinct phenomena can be observed, depending on salt concentration.
Screening of the electrostatic repulsion between the charged monomers by the ions onsets at low salt concentration ($\lesssim\SI{0.1}{\Molar}$). The polymer rapidly collapses from an extended coil ($\Rg \simeq \SI{4.7}{\nano\meter}$) to a more compact conformation as the salt is added and the Debye length, $\kappa^{-1} =(8\pi l_\mathrm{B} N_\mathrm{A} I)^{-1/2}$, decreases. Here, $N_\mathrm{A}$ is the Avogadro number and $I$ is molar ionic strength. This screening strengthens with the increasing salt concentration in the polymer phase. Weakly hydrated ions with large polymer affinity, that is, having greater $\epsca$, are more attracted to the polymer and cause stronger screening.

\begin{figure*}
	\centering
	\includegraphics[]{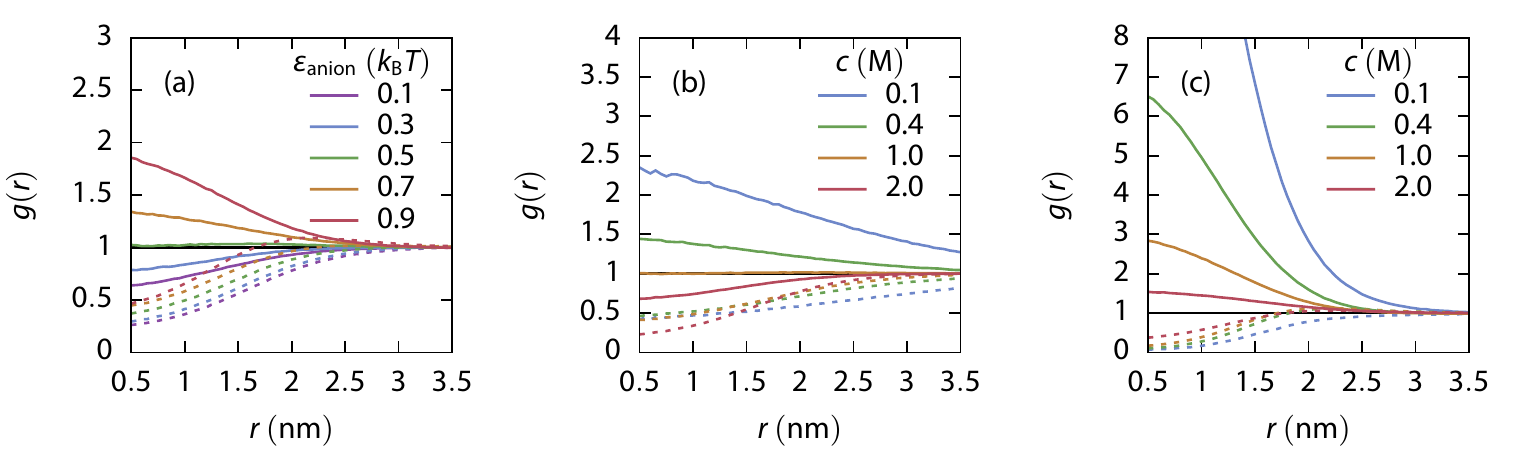}
	\caption[Radial distribution functions of ions around a center of mass of a cationic polymer. Snapshots of typical conformations of polymer chain under various conditions.]
	{Radial distribution functions of ions around a center of mass of a cationic polymer. Distributions of anions are shown in solid lines while the distribution of cations in dashed lines.
		(a) Various salt types at the isospheric point (a positively charged 200-mer bearing a charge fraction $\xi = 0.05$ in $\SI{1.0}{\Molar}$ salt solution). The cation is always strongly repulsive while the quality of the anion changes from repulsive ($\epsion = \SI{0.1}{\Boltzmann}$) to attractive ($\epsion = \SI{0.9}{\Boltzmann}$). 
		%(b) The effect of the charge fraction $\xi$ in $\SI{2.0}{\Molar}$ salt solution. with a repulsive ($\epsion = \SI{0.1}{\Boltzmann}$, solid lines), or an attractive ($\epsion = \SI{0.9}{\Boltzmann}$, dashed lines) anion. Radial distribution function of anions and corresponding cations are shown as thick and thin lines, respectively.
		(b) and (c) Effect of the salt concentration on the radial distribution function in a solution with positively charged polymer bearing a charge fraction $\xi = 0.10$. Salt with a repulsive ($\epsca = \SI{0.1}{\Boltzmann}$) and attractive ($\epsca = \SI{0.9}{\Boltzmann}$) anion is depicted in panel (b) and (c), respectively. 
	}
	\label{fig:sim-rdfs}
\end{figure*}

However, the collapse of the polymer chain beyond the size of the uncharged polymer in a salt-free solution, as in the case of $\epsca = \SI{0.9}{\Boltzmann}$, cannot be explained by the charge screening alone and another effect has to be taken into account. As the electroneutrality condition is fulfilled inside the polymer volume, the concentration of the strongly attractive anions is significantly higher than the concentration of the repulsive cations. The average $B_2$ of the ion-mer interactions is thus shifted towards more negative, \abbr{i.e.}, attractive, values, entering the bridging adhesion regime.\cite{Heyda2013} The radial distribution functions of anions and cations in \crefx{fig:sim-rdfs}{c} prove the greater abundance of strongly attractive anions compared to cations at low salt concentrations. Typical conformations of the polymer chain are shown in \crefx{fig:snapshots}{c,\,f} for the repulsive, and attractive salt, respectively. As salt concentration increases ($\gtrsim\SI{1}{\Molar}$), the salt specific effects begin to contribute to the interactions, resembling the case of the uncharged polymer, however with a notably different polymer size, cf. \crefx{fig:snapshots}{a--b,\,d--e}.

When the standard free energy of the coil-to-globule transition is treated as a function of salt concentration, it monotonically decreases for more repulsive ions with little affinity to the polymer. A different behavior is observed for the attractive ions: After an initial depression caused by the screening and bridging effects at low salt concentrations, the function rises because the swelling by attraction mechanism is introduced at high salt concentration. However, in the case of highly charged polymer chain ($\xi=0.10$), the screening-bridging mechanism dominates in the whole investigated concentration range (up to $\SI{2}{\Molar}$). Thus the salting-out power of ions is reversed when compared to the uncharged polymer.

The same simulation set is performed for a 200-mer bearing a lower charge fraction, $\xi=0.05$. Two salt concentration ranges can be distinguished based on the radius of gyration and the free energy of the coil-to-globule transition, \crefx{fig:sim-thermodynamics}{b,\,h}. When the salt concentration is less than $\SI{1}{\Molar}$, the screening and bridging mechanism dominates and the anions follow the reverse series regarding the salting-out power. At concentrations higher than $\SI{1}{\Molar}$, the specific ion effects are the driving force and the direct Hofmeister series of anions is obeyed.

The dividing point which separates the concentration ranges can be called the isospheric point as the radius of gyration of the polymer chain is the same regardless of the anion type. The isospheric point moves to the higher salt concentrations as the charge fraction of the polymer chain increases. For the uncharged polymer ($\xi=0.00$) it is trivially located at zero salt concentration, while for the highly charged polymer ($\xi=0.10$) at \SI{2}{\Molar} salt concentration. The $\Rg$ at the isospheric point corresponds approximately to the $\Rg$ of the uncharged polymer in a salt free solvent in our setup, \abbr{i.e.}, close to its radius of gyration under the critical conditions. 

Though the size of the polymer  at the isospheric point is the same, the concentration of the salt inside the polymer volume varies greatly depending on the salt type. Radial distribution functions of cations and anions around the polymer's center of mass at the isospheric point of the weakly charged polymer are shown in \crefx{fig:sim-rdfs}{a}. The concentration of the anions---and synergetically also of the cations---increases inside the effective polymer volume as the affinity of the anions to the polymer increases. The attractive anions even make the concentration of the highly repulsive cation higher around the polymer compared to the bulk solution.
%Despite the different salt concentrations, the mean value of the osmotic pressure of the salt inside the globule is the same at the isospheric point regardless of the salt type, as will be discussed later in \cref{sec:results-theory}.
Note, the salts with a highly attractive anion, however, do not intersect the isospheric point and the corresponding radius of gyration is achieved at a higher salt concentration, see \cref{si:fig:thermodynamics-high}. Very weakly hydrated ions, such as perchlorate \ce{ClO4-}, are experimentally known to deviate in a similar manner.\cite{Cho2009}

When we compare the data from simulation, \crefx{fig:sim-thermodynamics}{e}, to the experiments, \crefx{fig:experiment}{b}, a very good qualitative agreement can be found. When a small amount of salt is added, the LCST decreases rapidly in experiments and the order of anions follows the reverse series at low salt concentration, in accordance with the simulation predictions. Crossing the isospheric point at $\sim \SI{0.45}{\Molar}$, the anion order inverts and the direct Hofmeister series is recovered.

\subsection{Theoretical predictions}
\label{sec:results-theory}

For a better interpretation and categorization of the simulation results, we apply the mean-field theory (\cref{sec:theory}) for the very same systems. Minimization of the Gibbs free energy of the system provides the effective volume of the polymer in the thermodynamic equilibrium with the salt reservoir, or equivalently the effective radius $R_\mathrm{pol}$ if a spherical geometry is assumed. The polymers' effective radii in the thermodynamic equilibrium are depicted in \crefx{fig:theory-thermodynamics}{d--f} for an uncharged, weakly charged, and highly charged polymer chain, respectively. The results will not change significantly if a mean volume is used instead of the equilibrium volume.

The theory matches the simulation results almost quantitatively when the effective radius (\crefx{fig:sim-thermodynamics}{d--f}) and the radius of gyration, (\crefx{fig:theory-thermodynamics}{a--c}) are compared. The theory correctly predicts salting-out power of anions and its concentration dependence for the uncharged polymer. In the cases of weakly and highly charged polymer chains, the location of the isospheric points matches the simulation data very well. Below the isospheric point, the salting-out power of anions forms the reverse series, while above the direct series is obeyed, which is in agreement with both the simulations and the experimental data.

The osmotic pressure of the salt inside the effective volume of the polymer at the isospheric point is the same, regardless of the salt type, as can be deduced from the mechanical (pressure) equilibrium. The osmotic pressure of the reservoir depends only on the salt concentration in our model. The counter pressure originates from the polymer chain and the osmotic pressure of ions. The pressure of the polymer chain is determined by its spatial extension, thus it does not differ at the isospheric point. Hence the osmotic pressure of the ions inside the effective polymer volume has to be the same, independent of the salt type.
Note, however, the net salt-effect is a sum of two large contributions (see \cref{eq:fglob}), namely of an ideal gas and of an excess interaction, which significantly vary with the salt type (i.e., with $\epsca$).

\begin{figure}
	\centering
	\includegraphics[]{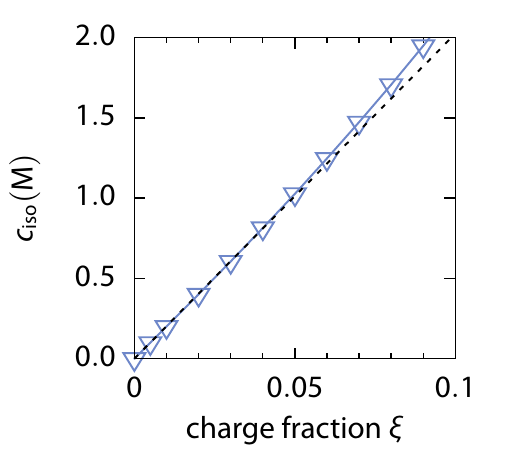}
	\caption[The isospheric point as a function of polymer's charge fraction $\xi$.]
	{The isospheric point as a function of the polymer charge fraction $\xi$ predicted by the mean-field theory. The isospheric point is computed as an intersection of $\Rg(\xi,c)$ curves of salts with $\epsca = \text{\SIlist{0.1;0.9}{\Boltzmann}}$, respectively (blue line with symbols). The dashed line shows linear asymptotic scaling at low charged fractions ($\xi \le 0.05$).}
	\label{fig:theory-isospheric}
\end{figure}

The salt concentration at the isospheric point $c_\mathrm{iso}$ scales linearly with the polymer's charge fraction $\xi$, provided the charge fraction is low ($\xi\lesssim 0.05$), see \cref{fig:theory-isospheric}. As the polymer's charge fraction increases, the isospheric point weakly deviates from the original linear trend and it is also gradually smeared out to a concentration range, as discussed in detail in the \refSI. Very weakly hydrated ions, \abbr{e.g.}, perchlorate \ce{ClO4-}, do not exhibit the isospheric point in experiments.\cite{Cho2009} Our theoretical model predicts such behavior for salts having $\epsca \gtrsim \SI{1.1}{\Boltzmann}$.

\begin{figure*}
	\centering
	\includegraphics[]{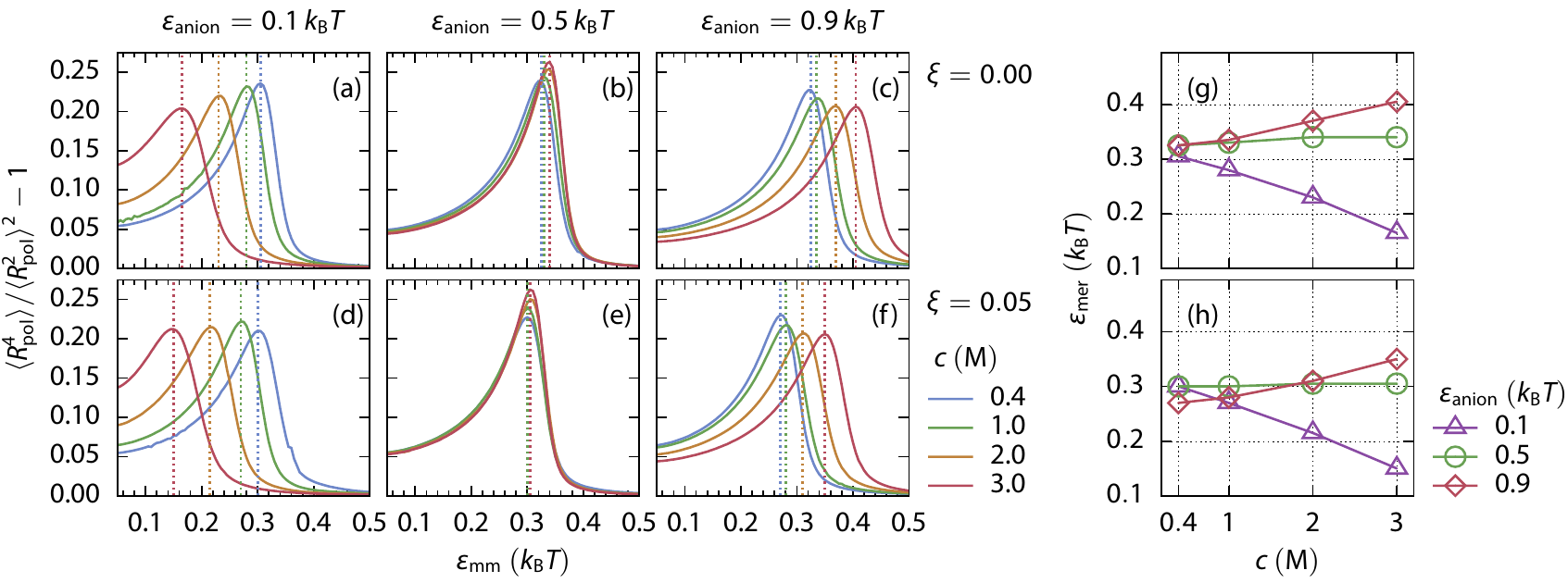}
	\caption[Fluctuations in the effective radius of the polymer chain $R\ipol$ as a function of the solvent quality at various salt concentrations, as derived from the theory.]
	{Fluctuations in the effective radius of the polymer chain $R\ipol$ as a function of the solvent quality at various salt concentrations (from \SIrange{0.4}{3.0}{\Molar}), as derived from the mean-field theory. The solvent quality is expressed as a strength of LJ interactions between mers, $\epsmer$. The top row (a--c) depicts the uncharged 200-mer ($\xi=0.00$), while the bottom row (d--f) shows fluctuations of the positively charged 200-mer with a charge fraction $\xi=0.05$. The left column (a, d) corresponds to a strongly hydrated (repulsive) salt ($\epsion = \SI{0.1}{\Boltzmann}$), the middle column (b, e) to a rather indifferent salt ($\epsion = \SI{0.5}{\Boltzmann}$), and the right column (c, f) to a weakly hydrated (attractive) salt ($\epsion = \SI{0.9}{\Boltzmann}$). The dashed lines indicate the maxima.}
	\label{fig:theory-fluctuations}
\end{figure*}

The fluctuation of the radius of gyration of a polymer chain is maximized under the critical conditions.\cite{Ivanov1998} The amount of fluctuations can be accessed from the theory using the Boltzmann distribution of the chain's states, \cref{eq:rg-boltzmann}. The dependence of the fluctuations on the solvent quality is depicted in \cref{fig:theory-fluctuations} for the uncharged and weakly charged polymer chains. With increasing concentration of repulsive salt (panels a,\,d), the maximum in fluctuations (\abbr{i.e.}, the critical conditions) is shifted to a better solvent quality (lower $\epsmer$). This corresponds to a lower LCST for ELPs. The opposite trend is observed for the attractive salt (panels c,\,f) where the solvent quality has to worsen to keep the critical conditions as the salt concentration increases. The relative shift of the critical solvent quality for the uncharged (c) and weakly charged (f) polymer is consistent with the difference in the standard free energy of the coil-to-globule transition obtained by simulations, \crefx{fig:sim-thermodynamics}{g,\,h}.

Let us stress again that the theory assumes a spherical geometry of the polymer and electroneutrality within this effective polymer volume. These assumptions are well fulfilled close to the critical conditions at $\gtrsim \SI{0.1}{\Molar}$ salt concentration. For example, our mean-field theory cannot predict the radius of gyration of a charged polymer without salt as no explicit electrostatic interaction between charged monomers is included. Also at low salt concentrations, the effective radius predicted by the theory deviates significantly from the radius of gyration. In such cases, different approaches have to be used.\cite{Khokhlov1980, Kundagrami2010, Budkov2015}

\subsection{Mean-field theory predictions for a negatively charged ELP}
\label{sec:results-anionic}

\begin{figure}
	\centering
	\includegraphics[]{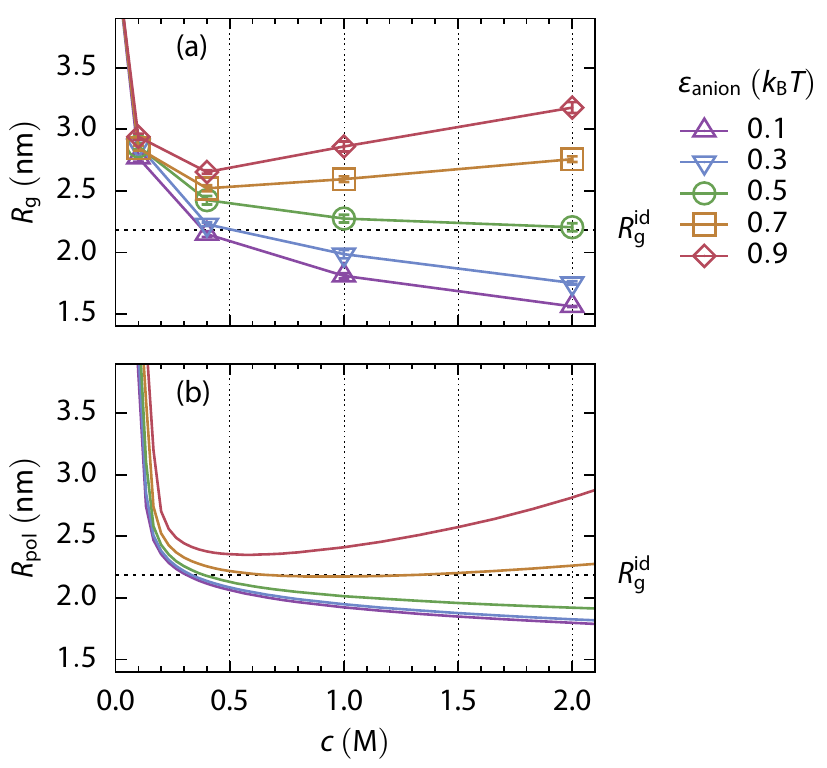}
	\caption{(a) Mean radius of gyration $\Rg$ of the negatively charged ELP (charge fraction $\xi=0.10$) as a function of salt concentration $c$ from the simulations. Salts with various anions and the same highly repulsive cation are used.
	%The quality of the anion $\epsca$ changes from repulsive (\SI{0.1}{\Boltzmann}) to attractive (\SI{0.9}{\Boltzmann}).
	Lines serve as guides to the eye. (b) A prediction of the effective radius of the polymer $R\ipol$ computed from the mean-field theory applied to the same systems.}
	\label{fig:elp-anionic}
\end{figure}

Having verified the descriptive power of the mean-field theory, we use it now to predict the effects of salts to a negatively charged polymer.  The set of salts remains the same, \abbr{i.e.}, the cation is highly repulsive while the type of the anion varies. The solvent quality has to worsen to $\epsmer = \SI{0.5}{\Boltzmann}$ to retain the critical conditions. The theory prediction is depicted in \crefx{fig:elp-anionic}{b}, while the simulation results which we use for verification are shown in panel (a) of the same figure. To the best of our knowledge, no systematic experimental data are yet available for such setup.

The simulation and theory results are in a very good mutual agreement. When the salt is added the collapse of the polymer chain is observed at first. The extend of the collapse is determined mainly by the (same) strongly hydrated cation and only mildly modulated by the anion type. Both electrostatic screening in the polymer and build-up of the osmotic pressure in the bulk solution contribute to the collapse. As the concentration of the salt increases, the anion-specific effects further drive the collapse or swelling, similarly to the case of an uncharged polymer. Contrary to the cationic polymer, neither the extent of the initial collapse dependents on the salt (anion) type, nor the isospheric point occurs. Consequently, the salting-out power of anions follows the direct Hofmeister series in the whole concentration range.

  \section{Conclusions}

The experimentally observed LCST response (and corresponding swelling and collapse) of elastin-like polypeptides (ELPs) to the addition of salts displays a rich behavior resulting from a combination of ion-specific screening of these weakly charged polyelectrolytes and direct ion-polymer interaction effects. We developed a minimalistic implicit-solvent CG model that can reproduce and qualitatively explain all the experimentally observed effects.\cite{Cho2008, Cho2009} Employing further simplifying assumption, we have developed a mean-field model which provides quantitative agreement with the CG computer simulations. Here, an effective second virial coefficient $B_2$ represents the quality of ions (strongly vs.\ weakly hydrated, or `kosmotropic' vs.\ `chaotropic' in classical terms) with respect to their affinity and preferential interaction with the polymer.

In particular, we demonstrated that for the uncharged chain ion-specific effects only are at work and the collapse is mediated through repulsion and swelling by attraction regimes; the salting-out power of ions follows a direct Hofmeister series characterized by a linear regime up to intermediate salt concentrations. For the charged, cationic polymers we demonstrated ion-specific screening and bridging at low salt concentration and strong ion-specific direct interaction effects at high salt concentration.
%Consistent with experiments, an isospheric point (charge fraction dependent) for the LCST vs.\ salt concentration was observed at which ion-specific effects vanish and the Hofmeister series inverts. That is, the salting-out power of ions follows a reverse series bellow the isospheric point and a direct one above. The same behavior is observed also experimentally in other positively charged proteins.\cite{Zhang2009}
We made an important theory-grounded observation of an isospheric point, \abbr{i.e.}, salt concentration at which ion-specific effects vanish and the Hofmeister series inverts. That is, the salting-out power of ions follows a reverse series bellow the isospheric point and a direct one above. The isospheric salt concentration depends on the polymer charge fraction. Our observation is supported by experimental data for ELPs\cite{Cho2008, Cho2009}, as well as for other positively charged proteins.\cite{Zhang2009}

For the anionic polymers our mean-field theory predicts a non-linear and for weakly hydrated (very attractive) anions also non-monotonic change of the LCST with salt concentration. At low salt concentrations, the strongly hydrated cation determines the screening and osmotic pressure in the polymer phase, only mildly modulated by the anion type. In the limit of high salt concentration, the ion-specific direct interactions drive the polymer collapse, or swelling, respectively, which results in two regimes of salt action. Consequently the direct Hofmeister series is followed in the whole concentration range, unlike for the cationic polymer.

Our work demonstrates the sensitivity of the structural and phase behavior of charged polymers to the addition of specific salts. Our mean-field model shall be useful for further extrapolation and guidance of related experiments.

  \renewcommand{\acknowledgementname}{\uppercase{Acknowledgement}}
  \begin{acknowledgement}
    \label{sec:acknowledgement}
The authors would like to thank Won Kyu Kim for a valuable discussion about the mean-field theory.
R.\,C. and J.\,D. acknowledge funding from the Deutsche Forschungsgemeinschaft (DFG grant DZ-74/6), Germany, for this project.  J.\,H. thanks the Czech Science Foundation (grant 16-24321Y) for support.

  \end{acknowledgement}

  \renewcommand{\suppinfoname}{\uppercase{Supporting Information Available}}
  \begin{suppinfo}
    Mathematica notebook implementing the mean-field theory; simulation setup details; trajectories and distributions of the radius of gyration; the isospheric point / highly attractive anions; evaluation of the Donnan potential.

  \end{suppinfo}

  \renewcommand{\refname}{\uppercase{References}}
  \bibliography{references}
\end{document}